\documentclass[twocolumn,amssymb,nobibnotes,aps,showpacs,pra]{revtex4}
  \usepackage[dvips]{color}
  \usepackage{newlfont}
  \usepackage{graphicx}
  \usepackage{amssymb}
  \usepackage{amsmath}
  \usepackage[latin1]{inputenc}
  \usepackage{float}

\begin{document}
\title{Decoherence induced by a phase-damping reservoir}
\author{Renato M. Angelo}
\affiliation{Departamento de Física, Universidade Federal do Paraná, Caixa Postal 19044, CEP 81531-990, Curitiba, Paraná, Brazil}
\author{E. S. Cardoso}
\author{K. Furuya}
\affiliation{Instituto de Física `Gleb Wataghin', Universidade Estadual de Campinas, Caixa Postal 6165, CEP 13083-970, Campinas, São Paulo, Brazil}

\begin{abstract}
A phase damping reservoir composed by $N$-bosons coupled to a system of interest through a cross-Kerr interaction is proposed and its effects on quantum superpositions are investigated. By means of analytical calculations we show that: i-) the reservoir induces a Gaussian decay of quantum coherences, and ii-) the inherent incommensurate character of the spectral distribution yields irreversibility. A state-independent decoherence time and a master equation are both derived analytically. These results, which have been extended for the thermodynamic limit, show that nondissipative decoherence can be suitably contemplated within the EID approach. Finally, it is shown that the same mechanism yielding decoherence are also responsible for inducing dynamical disentanglement.


\end{abstract}

\maketitle

\section{Introduction}

Environment-induced decoherence (EID) has been widely accepted as
the mechanism responsible for the destruction of quantum
superpositions. A great amount of research in the last two decades
has put in evidence - theoretically \cite{zurek} and
experimentally \cite{joos} - the potentiality of this effect in
explaining challenging issues in quantum mechanics, as for
instance the nonexistence of macroscopic superpositions and the
measurement process. The EID approach is based on the coupling of
a system $S$ with a many degrees of freedom environment, which is
theoretically described in terms of microscopic quantum particles
in equilibrium at temperature $T$. The coupling with the reservoir
forces the system to evolve through a nonunitary, irreversible
and, in general, dissipative dynamics.

One of the conceptual difficulties of the EID program, as it has
been pointed out by some authors, concerns the irreversible energy
 transfer from the system to the reservoir, which prohibits the
application of this approach to conservative systems. Several
alternative methods have been proposed to settle this problem. In
general, these approaches defend the so termed {\em self-induced
decoherence} (SID), whose cause is not attributed to external
degrees of freedom (environment). Among them one may quote, for
instance, proposals of a generalized Schrödinger equation
\cite{milburn}, assumptions about fluctuations of classical
parameters or internal degrees of freedom yielding decoherence
\cite{bonifacio}, and nondissipative decoherence as a result of
the quantum state averaging over a thermal distribution of
velocities \cite{ford}. Also, references \cite{frasca,cast97}
contribute to this list. 

A recurrent result in the SID program is the reduction of the system dynamics to the following form of the well-known phase-destroying master equation \cite{walls,gardiner}:
\begin{eqnarray}
\frac{\partial \widehat{\rho}}{\partial t}=\frac{[\,\widehat{H},\widehat{\rho}\, ]}{\imath\,\hbar}-\frac{\tau}{\hbar^2}\,[\,\widehat{H},[\,\widehat{H},\widehat{\rho}\,]\,],\label{PDME}
\end{eqnarray}
$\tau$ being a specific parameter in each theory. By means of straightforward calculations in the eigenbasis $\{|E_n\rangle\}$ of $\widehat{H}$ one may prove that the above master equation admits the following solution:
\begin{eqnarray}
\widehat{\rho}(t)&=&\sum\limits_{n,n'=0}^{\infty} \rho_{n n'}\,e^{-\imath\,(E_n-E_{n'})t/\hbar} \times \nonumber \\ && \exp\left[-\frac{\gamma(t)\,(E_n-E_n')^2}{\hbar^2}\right] \,|E_n\rangle\langle E_{n'}|,
\end{eqnarray}
where $\rho_{n n'}=\langle E_n|\widehat{\rho}(0)|E_n'\rangle$ and $\dot{\gamma}=\tau$. The purity, $\cal{P}=\textrm{Tr}[\rho^2]$, then reads
\begin{eqnarray}
\cal{P}(t)=\sum\limits_{n,n'=0}^{\infty} |\rho_{n n'}|^2\, \exp\left[-\frac{2\,\gamma(t)\,(E_n-E_n')^2}{\hbar^2}\right].\label{Psid}
\end{eqnarray}
When $\tau$ is constant, for example, one can observe an exponential decay with time for the off-diagonal terms of $\widehat{\rho}$, implying in a decrease of the purity $(\cal{P}(0)>\cal{P}(\infty))$, even though the energy of the system remain unchanged, i.e., $\textrm{Tr}[\widehat{H}\,\widehat{\rho}(t)]=\textrm{Tr}[\widehat{H}\,\widehat{\rho}(0)]$. Clearly, the master equation \eqref{PDME} suitably describes nondissipative phase decoherence. Recently, however, the relevance of SID as a physically meaningful process has been questioned \cite{max}.

In this work we show that it is also possible to obtain phase
decoherence within the EID scenario, without any additional
assumption to the standard quantum formalism if we consider a diagonal interaction. Our approach is
based on a many-boson model in which the interaction between the
components is given by the well-known cross-Kerr coupling
$\widehat{a}^{\dag}_i\,\widehat{a}_i\,\widehat{a}^{\dag}_j\,\widehat{a}_j$.
This assumption is grounded in the fact that this kind of interaction is physically conceivable, i.e., two hybrid bosonic degrees of freedom such as ionic vibrational mode and the quantized field may interact effectively in this way as has been demonstrated recently \cite{fernando}. Besides, several information transfer processes have been developed based on cross-Kerr interactions \cite{lee} in such hybrid systems, and several sources of decoherence without transition between the states of different quantum numbers - thus destroying coherence without causing energy dissipation - have also been discussed in \cite{murao}. These facts lead us to think that decoherence under this kind of phase channel will be a real problem to be considered. One can also do engineering of a bath of $N$ identical ion traps \cite{myatt} each one coupled to the same field, such that they satisfy the following conditions of Ref. \cite{fernando}: (i) the detuning of the atom with the field $\Delta$ is different from the multiples of the frequency $\nu$ of the vibrational mode ($\Delta \ne k \nu$, $k$ integer) thus avoiding to excite the vibrational degree of freedom of each ion; (ii) in the large detuning limit ($g \ll \Delta$; $g$ being the ion-field coupling) in order to achieve a dispersive regime; and (iii) $\Delta \ll \nu$ if the non-Lamb-Dicke regime is desirable. Although usually the energy dissipating sources of decoherence may also be present (via typical oscillator-oscillator position coupling), in the above engineered bath case, a nondissipative kind of decoherence is expected. By means of analytical calculations we show that the model indeed produces nondissipative decoherence, transforming quantum superpositions into mixed states through a Gaussian decay.

   In addition, we show that bipartite entanglement can be
dynamically destroyed by such a reservoir. Dodd and Halliwell
showed how open system dynamics destroys entanglement by means of
the same mechanism that destroys interference \cite{dodd}. Here,
analytical formulas allow us to make the same identification
rigorously.

This paper is outlined as follows. In Sec. \ref{mod} the reservoir is modeled as a system of $N$ harmonic oscillators in a thermal equilibrium and the coupling with the system of interest is assumed to be given effectively by a cross-Kerr interaction. Exact results demonstrating the capability of the model in yielding decoherence for arbitrary quantum states are presented in Sec.\ref{decoherence}, where: (A) analytical results demonstrating the Gaussian decay of quantum coherences and an estimate for the decoherence time are obtained for short times; (B) the equilibrium regime is analyzed; (C) the thermodynamic limit is implemented; (D) two examples of decohering dynamics for some special quantum states, such as the cat state and a superposition of Fock states, are given, and (E) a phase-destroying master equation is derived in the regime of weak coupling. In Sec.\ref{disent} we demonstrate the dynamical disentanglement induced by the same mechanisms responsible for phase decoherence. Finally, in Sec.\ref{conclusions} we relate our concluding remarks.

\section{The Model}\label{mod}

In ref.\cite{fernando}, Semião and Barranco have derived an effective Hamiltonian for a model of two-level trapped ion coupled to the quantized field inside a cavity. Under certain conditions, which allow for the rotating wave approximation and assumptions on both a large detuning between the ion and the field (dispersive limit) and a small value for the Lamb-Dicke parameter, the authors find the following effective Hamiltonian
\begin{eqnarray}
\widehat{H}_{eff}=\Omega\,\widehat{a}^{\dag}_1\,\widehat{a}_1+\omega\,\widehat{a}^{\dag}_2\,\widehat{a}_2+\lambda\,\widehat{a}^{\dag}_1\,\widehat{a}_1\,\widehat{a}^{\dag}_2\,\widehat{a}_2.
\label{Heff}
\end{eqnarray}
This Kerr-type coupling produces only phase interaction, not being able to introduce energy exchanges between the bosons. For this reason, this kind of phase coupling turns out to be a potential candidate for modeling nondissipative reservoirs.

Having been inspired by \eqref{Heff}, we consider the  following many-bosons model
\begin{eqnarray}
\widehat{H}=\sum\limits_{i=0}^N \hbar\,\omega_i\,\widehat{n}_i+\sum\limits_{i,j=0 \atop i>j}^N \hbar\,g_{ij}\,\widehat{n}_i\,\widehat{n}_j,
\label{model}
\end{eqnarray}
where $\widehat{n}_i$ denotes the number operator of the $i$-th boson. The coupling is clearly incapable to produce energy exchanges between any two degrees of freedom.

Now we focus on the dynamics of a single boson, say the one with index ``0'', in order to investigate the features of such an environment. For simplicity, we also assume that there is no interaction between the reservoir components.  The Hamiltonian \eqref{model} is then replaced by
\begin{eqnarray}
\widehat{H}&=&\widehat{H}_S+\widehat{H}_R+\widehat{H}_I,\label{H}
\end{eqnarray}
where
\begin{subequations}
\begin{eqnarray}
\widehat{H}_S&=&\hbar\,\omega_0\,\widehat{n}_0, \label{H0}\\
\widehat{H}_R&=&\sum\limits_{k=1}^N\hbar\,\omega_k\,\widehat{n}_k, \label{HR}\\
\widehat{H}_I&=&\sum\limits_{k=1}^N \hbar\,g_{0k}\, \widehat{n}_0\,\widehat{n}_k.\label{HI}
\end{eqnarray}
\end{subequations}
Initially, the joint quantum state is assumed to be
\begin{eqnarray}
\widehat{\rho}(0)=\widehat{\rho}_S(0)\otimes\frac{e^{-\beta\widehat{H}_R}}{\textrm{Tr}\left[e^{-\beta\widehat{H}_R}\right]},
\end{eqnarray}
which indicates that all oscillators of the reservoir are in thermal equilibrium at a temperature $(k_B\beta)^{-1}$. The system of interest is initially in an arbitrary pure state $\widehat{\rho}_S(0)$.

 As it has been shown by Ford {\it et al} \cite{ford88}, the most general quantum Langevin equation, describing macroscopically the dissipative dynamics of a quantum system, can be realized with an independent oscillator (IO) model. Also, these authors have demonstrated that several traditional models of reservoirs inducing dissipative decoherence can be mapped on an IO model, which, accordingly, may be regarded as a fundamental model for the EID program.

Our reservoir, however, may be regarded as being grounded on a type of generalized IO model which implements nonlinear couplings. In fact, the original model for the ions, from which the effective cross-Kerr coupling given in Eq.\eqref{Heff} has been derived in ref.\cite{fernando}, contains the fundamental ingredients of an IO model, including energy exchanges.

\section{Phase Decoherence}\label{decoherence}
The rather simple form of the coupling allows one to solve
analytically the dynamics by means of straightforward
calculations, with no prior need for weak coupling or specific
statistical assumptions. The exact result for the system density
matrix  in the $\widehat{H}_0$ eigenbasis is written as
\begin{eqnarray}
\widehat{\rho}_S(t)=\hspace{-0.3cm}\sum_{m,m'=0}^{\infty} \hspace{-0.2cm}\rho_{m,m'}\,C_{m,m'}(t)\,e^{-\imath\,(m-m')\,\omega_0 t}\,|m\rangle\langle m'|,\label{rhoSt}
\end{eqnarray}
where $\widehat{\rho}_S=\textrm{Tr}_R[\,\widehat{\rho}\,]$, $\rho_{m,m'}=\langle m|\widehat{\rho}_S(0)|m'\rangle$ and
\begin{eqnarray}
C_{m,m'}(t)=\prod\limits_{k=1}^N\left[\frac{1-e^{-\beta\,\hbar\,\omega_k}}{1-e^{-\beta\,\hbar\,\omega_k}\,e^{-\imath\,(m-m')\,g_{0k}\,t}}\right].\label{C}
\end{eqnarray}
The purity of the system state, $\cal{P}_S= \textrm{Tr}_S[{\widehat{\rho}_S}²]$, is given by
\begin{eqnarray}
\cal{P}_S(t)=\sum\limits_{m,m'=0}^{\infty}  \left|\rho_{m m'} \right|^2\,\left|C_{m,m'}(t)\right|^2,
\label{PS}
\end{eqnarray}
with
\begin{eqnarray}
\left|C_{m,m'}(t) \right|^2=\prod\limits_{k=1}^N\left\{ 1+
\frac{\sin^2\left[\frac{(m-m')\,g_{0k}\,t}{2}
\right]}{\sinh^2\left[\frac{\beta\,\hbar\,\omega_k}{2}
\right]}\right\}^{-1}, \label{C2}
\end{eqnarray}

Notice in Eq.\eqref{PS} that all the time and temperature
dependences come from the function $C_{m,m'}$. Also, in
\eqref{rhoSt} we see that $C_{m,m'}$ is the only function
depending on the coupling parameters between the system and the
reservoir. Thus, it is natural to expect that this function
carries all the dynamical informations about the correlations of
the system with the environment and, therefore, about the
decoherence of the quantum state.

\begin{figure}[ht]
\begin{center}
\includegraphics[scale=0.35,angle=-90]{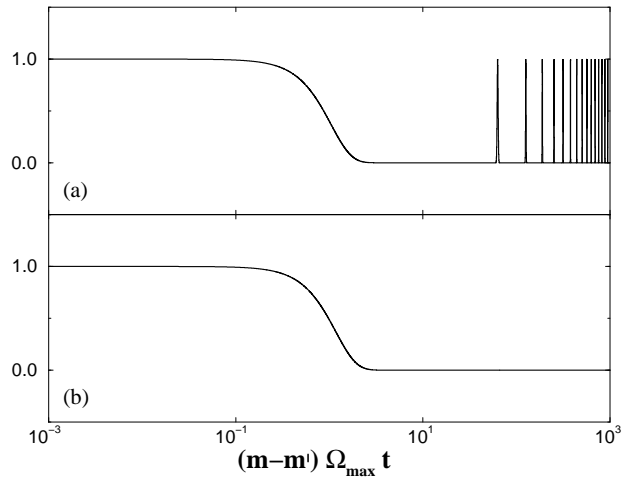}
\end{center}
\caption{Log-linear plot for the function $|C_{m,m'}|^2$ as a function of $(m-m')\,\Omega\, t$ for $N=100$ and $\beta\,\hbar\,\omega=1$. Two cases of couplings were simulated, namely: (a) a resonant spectrum, $g_{0k}=\Omega/\sqrt{N}$, and (b) a Gaussian spectral distribution, $g_{0k}=\left(\Omega/\sqrt{N}\right)\,\exp\left[-\frac{\pi\,(k-k_0)^2}{2\,N^2}\right]$. In (a) $\Omega=0.1$ and in (b) $\Omega=0.1125$. In both cases, $\cal{G}^2\equiv \sum_k g_{0k}^2=0.01$ (see Eq.\eqref{tauD}) and $k_0=50$. Notice the long-time recurrences in (a).} \label{fig1}
\end{figure}

The analytical results \eqref{rhoSt}-\eqref{C2} provide a rather
interesting explanation for the well-known requirement of a
spectral distribution for the coupling parameters $g_{0k}$. For
resonant ($g_{0k}=g$) or commensurate ($g_{0k}= l_k \,g$, $l_k$
integer) couplings, the time dependence in Eq.\eqref{C} behaves as
$e^{-\imath M g\, t}$, with $M$ integer. Then, for all instants
satisfying $t_n=2\pi n/g$, with $n\in \mathbb{N}$, one has
$C_{m,m'}=1$, for all $m$ and $m'$, indicating that the quantum
state has just recovered its initial condition of purity, namely
$\cal{P}_S(t_n)=\cal{P}_S(0)=1$. This certainly is an unexpected
recurrence for a supposedly irreversible reservoir. However, far
apart from this very especial conditions, incommensurate values of
coupling inherent to any physically acceptable spectral
distribution will set irreversibility just by avoiding
simultaneous recurrences in $C_{m,m'}$. This is indeed ensured by
the product in \eqref{C}. These statements are demonstrated in
Fig.\ref{fig1}, where a typical behavior for $|C_{m,m'}|^2$ is
numerically simulated.
Notice that all arguments given in this section remain valid even
for the simplest case of resonant harmonic frequencies
($\omega_k=\omega$). Since this assumption does not change significantly
the main features of the model, it will be considered hereafter for simplicity.

\subsection{The short-time regime}

In Fig.\eqref{fig1} one may clearly realize the presence of a
short-time regime in which the collapse of the characteristic
function occurs. Actually, in this section we shall show
analytically that this is the very time scale of decoherence.

Let us consider the regime of short times, namely $(m-m')\, g_{0k}\,
t\ll 1$. In this limit Eq.\eqref{C2} can be replaced with
\begin{eqnarray}
\left|C_{m,m'}(t)\right|^2 \cong
e^{-(m-m')^2\,t^2/\tau_{D}^2},\label{decay}
\end{eqnarray}
where the decoherence time $\tau_D$ is given by
\begin{eqnarray}
\tau_{D}=2\,\cal{G}^{-1}\,\sinh\left(\beta\,\hbar\,\omega/2\right).\label{tauD}
\end{eqnarray}
The frequency $\cal{G}$, defined by $\cal{G}^2\equiv \sum_k g_{0k}^2$, does not depend on the specific profile of the spectral distribution, since it contains the information only about the global sum, which is supposed to be normalized (the reason for that will be discussed latter). Indeed, in Eq.\eqref{decay} (and in Fig.\ref{fig1}) one may see that the decay occurs even for a resonant reservoir.  Then, one may conclude that the only role played by the spectral distribution is to provide the character of irreversibility for the reservoir. 

 Note, by Eq.\eqref{decay}, that the decoherence time has been defined regardless of the difference $m-m'$, even though the instant of decay is strongly dependent on this difference. This choice attributes to $\tau_D$ the role of an upper bound for surviving coherences, i.e., it corresponds to the time scale beyond which decoherence has indeed affected all elements of the matrix $\widehat{\rho}_S$.

Another important characteristic of our model is the Gaussian
decay given by Eq.\eqref{decay}, which has already been found in
other approaches \cite{ford, bonifacio}. This behavior implies in
a sudden decrease in the purity of the (arbitrary) quantum state
for times longer than $\tau_{D}$. In fact, well beyond this time
scale the quantum state will have reached its equilibrium
configuration. In this regime, since
$C_{m,m'}(t)=|C_{m,m'}(t)|\,e^{\imath\,\Phi(t)}$, one may write
\begin{eqnarray}
C_{m,m'}(t> \tau_D)=\delta_{m,m'},
\end{eqnarray}
indicating that the dynamics will transform
\begin{eqnarray}
\widehat{\rho}_S(0)=\sum_{m,m'=0}^{\infty}
\rho_{m,m'}\,|m\rangle\langle m'|, \label{rho0}
\end{eqnarray}
with
\begin{eqnarray}
\cal{P}(0)=\sum_m \rho_{m,m}^2+\sum\limits_{m,m' \atop m\neq m'}\left|\rho_{m,m'}\right|^2,\label{P0}
\end{eqnarray}
into
\begin{eqnarray}
\widehat{\rho}_S(t>\tau_D)=\sum_{m=0}^{\infty}
\rho_{m,m}\,|m\rangle\langle m|, \label{rho8}
\end{eqnarray}
with
\begin{eqnarray}
\cal{P}(t>\tau_D)=\sum_m \rho_{m,m}^2.\label{Ptau}
\end{eqnarray}

Note that the structure of the coupling given by Eq.\eqref{HI}
automatically determines the pointer basis, i.e., the basis in
which the density matrix becomes diagonal. This allows us to
conjecture that this reservoir will fatally destroy the
entanglement for a state like $|n\rangle|m\rangle+|n'\rangle|m'\rangle$
(this will be demonstrated in Sec.\ref{disent}). 

 Another remarkable result derives from equations \eqref{P0} and \eqref{Ptau}: $\cal{P}(0)$ is always greater than $\cal{P}(t>\tau_D)$, indicating the occurrence of a base-independent decoherence. Furthermore, since no assumption was made about the pure initial state, the decohering process is also state-independent, $\tau_D$ being the general upper bound for the time scale of such a process.

 This generality associated to $\tau_D$ attests that this time scale is in fact an intrinsic feature of the reservoir model and not of the particular quantum state of the system. This is perhaps the most important difference regarding the usual time scales found in literature, which in general exhibit a dependence with $d^{-n}$, $d$ being the distance in the configuration space separating the two components of the particular superposition state under investigation (for examples of such a state-dependent time scales see references \cite{zurek,joos}, for the free particle problem, and \cite{connell03,ford04}, for oscillators models). In our model this dependence might be introduced by incorporating the ``distance'' $m-m'$ in the definition given by Eq.\eqref{tauD}. However, we have opted to attribute a general character to $\tau_D$. Concerning the dependence with other physical parameters, such as characteristic frequencies, temperature and $\hbar$, our time scale agrees qualitatively with those of the references quoted above. For instance, all time scales predict a decrease in the decoherence time as the temperature increases. A precise quantitative agreement is not expected in principle since we are working, in contrast to those approaches, with nonlinear interactions.

\subsection{The long-time regime}

\begin{figure}[ht]
\begin{center}
\includegraphics[scale=0.35,angle=-90]{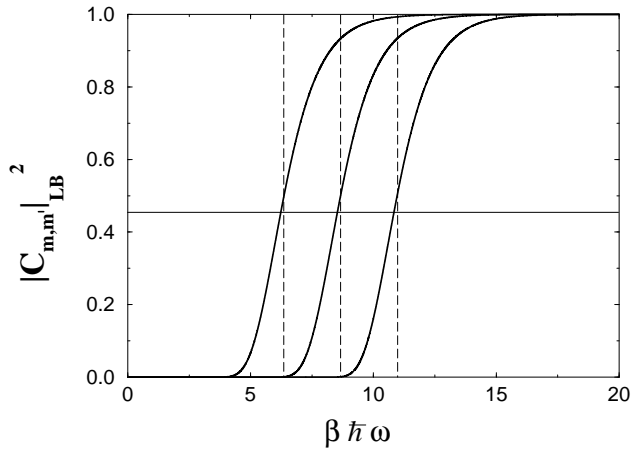}
\end{center}
\caption{\small Lower bound $|C|_{LB}^2$ as a function of the dimensionless parameter $\beta\,\hbar\,\omega$ for $N=10^2$, $10^{3}$, and $10^{4}$ (from left to right). The critical values $\beta_{crit}\,\hbar\,\omega$, estimated from Eq.\eqref{betaC} for each $N$, are given by the vertical dashed lines. }
\label{fig2}
\end{figure}

Now we analyze the equilibrium situation $(t\to \infty)$. In this
case, a direct inspection of Eq.\eqref{C2} allows us to calculate
a {\it lower bound}, which for any $m$ and $m'$, reads
\begin{eqnarray}
\left|C_{m,m'}\right|^2_{LB} =
\left[\frac{\sinh^2\left(\beta\,\hbar\,\omega/2
\right)}{1+\sinh^2\left(\beta\,\hbar\,\omega/2
\right)}\right]^N.\label{LB}
\end{eqnarray}
It is easy to see, for $N$ large but finite, that the lower bound given by Eq.\eqref{LB} can range from arbitrarily small values, in the regime of high temperatures ($\beta\,\hbar\,\omega\ll 1$), to values arbitrarily close to the unit, in the regime of low temperatures ($\beta\,\hbar\,\omega\gg 1$). The latter regime, for which $|C_{m,m'}|^2_{LB}\to 1$, corresponds to the one in which the dynamics is practically unitary, without decoherence. On the other hand, the former, for which $|C_{m,m'}|^2_{LB}\to 0$, corresponds to the regime in which the effects of decoherence are accentuated. In Fig.\ref{fig2} these regimes can be identified clearly.

The abrupt transition between these two limits (central region of Fig.\ref{fig2}) corresponds to a regime in which decoherences still occurs, but only partially, since a finite value for $|C_{m,m'}|^2_{LB}$ indicates the existence of surviving off-diagonal terms in the quantum state. The critical temperature, i.e., the one beyond which the effects of decoherence become accentuated, may be estimated by imposing that $|C_{m,m'}|^2_{LB}=1/2$ in Eq.\eqref{LB}. The result is a function of $N$:
\begin{eqnarray}
\beta_{crit}(N)=\left(\frac{\hbar\,\omega}{2}\right)^{-1}\,\textrm{arcsinh}\left(\frac{1}{\sqrt{2^{1/N}-1}}
\right).\label{betaC}
\end{eqnarray}
For temperatures in the vicinity of $(k_B\beta_{crit})^{-1}$ the quantum
state must contain off-diagonal terms contributing to an increase
in the value of the purity in the regime of long times.

\subsection{The  thermodynamic limit}\label{thermo}

In the previous sections we have discussed the general characteristic of our cross-Kerr reservoir for an arbitrary value of $N$. Now, we investigate the thermodynamic limit, which better describes the physical reality of a reservoir.

Rigorously, the thermodynamic limit is implemented by taking $N\to\infty$ and $V\to\infty$, with $(N/V)\to 1$. Before considering this limit in our model, however, it would be necessary to replace the coupling $g_{0k}$ by its original physical form $g_{0k}/\sqrt{V}$, which derives from the process of quantization of the electromagnetic field \cite{scully}. With this correction, the frequency $\cal{G}$ defined in Eq.\eqref{tauD} would assume the form $\cal{G}=\sqrt{\sum_k g_{0k}^2/V}$. Then, in the resonance $g_{0k}=\Omega$, for instance, we would have $\cal{G}=\Omega \sqrt{N/V}$, which possesses a well defined thermodynamic limit $\Omega$. 

This simple example suggests that we can implement the thermodynamic limit in our formulas just by requiring that the coupling $g_{0k}$ be quadratically normalized, i.e., that $\cal{G}^2=\sum_k g_{0k}^2$ be finite. For instance, for the resonant case mentioned above, the assumption that $g_{0k}=\Omega/\sqrt{N}$ will automatically set $\cal{G}=\Omega$. In a less trivial case, in which the coupling is given by a Gaussian spectral distribution, the thermodynamic limit can be ensured just by assuming that
\begin{eqnarray}
g_{0k}=\frac{\Omega}{\sqrt{N}}\exp\left[-\frac{\pi\,(k-k_0)^2}{2\,N^2}\right].
\label{g0k}
\end{eqnarray}
In this case, one may verify numerically that the relation $\cal{G}< \Omega$ is always obeyed for any $N$. For instance, for $k_0=50$, one may verify that $\cal{G}\to 0.706\,\Omega$ as $N\to\infty$. Also in this case, it may be verified that $\cal{G}\leqslant 0.889\,\Omega$, this upper bound occurring for $N=2 k_0=100$. 

With these considerations we can reinterpret the results given by Esq.\eqref{decay}-\eqref{betaC} within the framework of a reasonable thermodynamic reservoir possessing the following characteristics:
\begin{itemize}
\item[i-)] Gaussian decay for the short-time regime;
\item[ii-)] Decoherence time given by $\tau_D=2\,\cal{G}^{-1}\,\sinh\left(\beta\,\hbar\,\omega\right)$;
\item[iii-)] Characteristic frequency given by $\cal{G}^2=\sum\limits_{k=1}^{\infty}g_{0k}^2$;
\item[iv-)] Irreversibility induced by a quadratically normalizable spectral distribution ($g_{0k}$);
\item[v-)] Lower bound $|C|²_{LB}=0$ and $T_{crit}=0$.
\end{itemize}
Item v-), which can be directly obtained by taking $N\to\infty$ in Eq.\eqref{LB} and Eq.\eqref{betaC}, tell us that decoherence will occur in the thermodynamic limit even for arbitrarily low temperatures.

\subsection{Examples}\label{applications}

\begin{figure}[ht]
\begin{center}
\includegraphics[scale=0.35,angle=-90]{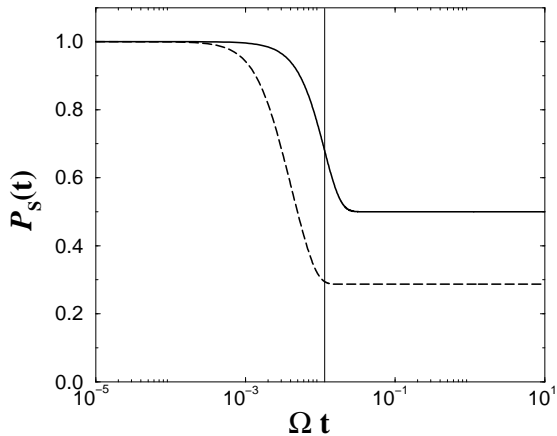}
\end{center}
\caption{\small Log-linear plot for the purity of (i) a cat state
(dashed curve), with $\alpha=2.0$, and (ii) the superposition
$(|1\rangle+|2\rangle)/\sqrt{2}$ (solid curve). For these
simulations, $N=100$ and $\beta\,\hbar\,\omega=0.01$. Here we have
used the spectral distribution employed in Fig.\ref{fig1}-(b). The
vertical line stands for the decoherence time $\Omega\,\tau
_D=1.13\times 10^{-2}$, estimated from Eq.\eqref{tauD}.}
\label{fig3}
\end{figure}

Any initially pure state for the system of interest may be expanded as
\begin{eqnarray}
|\psi_0\rangle=\sum\limits_{n=0}^{\infty} p(n)\,|n\rangle,
\end{eqnarray}
where $|p(n)|^2$ is the photon distribution defining the quantum state. The corresponding density operator is then written as $\widehat{\rho}_S(0)=|\psi_0\rangle\langle\psi_0|$ and its matrix elements in Fock basis reads $\rho_{m,m'}=p(m)\,p^{*}(m')$. Thus, Eq.\eqref{PS} becomes
\begin{eqnarray}
\cal{P}_S(t)=\sum\limits_{m=0}^{\infty}\sum\limits_{m'=0}^{\infty}\left|p(m)\right|^2\,\left|p(m')\right|^2\,\left|C_{m,m'}(t)\right|^2.\label{PS1}
\end{eqnarray}
Now we apply our model to investigate numerically the decoherence process for: (i) the cat state
\begin{eqnarray}
|\psi_0\rangle=\frac{|\alpha\rangle+|-\alpha\rangle}{\sqrt{2\left(1+e^{-2|\alpha|^2} \right)}}
\end{eqnarray}
 and (ii) the superposition of Fock states
\begin{eqnarray}
|\psi_0\rangle=\frac{|m_1\rangle+|m_2\rangle}{\sqrt{2}},
\end{eqnarray}
with $m_1\ne m_2$. For this Fock superposition, for which an analytical calculation yields $P_S(t)=\left(1+|C_{m_1,m_2}(t)|^2 \right)/2$, it is easy to verify the classicalization just by applying formulas \eqref{rho0} and \eqref{rho8}. Indeed, the equilibrium value for the purity, $\cal{P}_S(\infty)=1/2$, is directly associated to the statistical mixture $\widehat{\rho}_S(\infty)=\left(|m_1\rangle\langle m_1|+|m_2\rangle\langle m_2| \right)/2$. For the cat state, on the other hand, at the equilibrium time we have a mixed state with
$0<\cal{P}_S(\infty)<1$. In Fig.\ref{fig3} we show numerical
simulations for the purity of these states. The decoherence time
is also plotted, attesting both the validity and universality of
the estimate \eqref{tauD}. Notice that for $\Omega\simeq 10^{6}$
s$^{-1}$ we have $\tau_D\simeq 10^{-8}$ s.

\subsection{Master Equation}\label{secME}

The interaction Hamiltonian given by Eq.\eqref{HI} can be written as
\begin{eqnarray}
\widehat{H}_I= \widehat{V}_S \otimes \widehat{V}_R,
\end{eqnarray}
where we have defined: $\widehat{V}_S=\widehat{n}_0$ and $\widehat{V}_R=\sum_k g_{0k}\,\widehat{n}_k$. By means of relatively simple calculations we then deduce a master equation for our model in the weak coupling approximation. The result, in the interaction picture, reads
\begin{eqnarray}
\frac{\partial \widehat{\rho}_S^I}{\partial t}=\frac{\left[\widehat{\mathbb{V}}_S,\widehat{\rho}_S^I \right]}{\imath\,\hbar}-\frac{\xi\,t}{\hbar^2}\,\left[\widehat{\mathbb{V}}_{S},\left[\widehat{\mathbb{V}}_S,\widehat{\rho}_S^I \right] \right],\label{ME}
\end{eqnarray}
where $\widehat{\mathbb{V}}_S\equiv \langle \widehat{V}_R \rangle\,\widehat{V}_S$ and $\xi\equiv \langle \widehat{V}_R^2\rangle/\langle \widehat{V}_R\rangle^2$. The reservoir expectations, $\langle\widehat{V}_R^n\rangle=\textrm{Tr}\left[\widehat{\rho}_R(0)\,\widehat{V}_R^n\right]$, are given exactly by
\begin{subequations}
\begin{eqnarray}
\langle \widehat{V}_R\rangle&=&\sum_{k=1}^N
\frac{g_{0k}}{e^{\beta\,\hbar\,\omega}-1}, \\
\langle\widehat{V}_R^2\rangle&=&\sum_{k=1}^N
g_{0k}^2\,\frac{e^{\beta\,\hbar\,\omega}+1}{\left(e^{\beta\,\hbar\,\omega}-1
\right)^2}+\langle \widehat{V}_R\rangle^2.
\end{eqnarray}\label{<V>}
\end{subequations}
Using the spectral distribution given by Eq.\eqref{g0k} one can rewrite these formulas as
\begin{subequations}
\begin{eqnarray}
\langle \widehat{V}_R\rangle&=&\Omega\,\frac{\sqrt{2\,N}}{e^{\beta\,\hbar\,\omega}-1}\,A_2(N), \\
\langle\widehat{V}_R^2\rangle&=&\Omega^2\,\frac{e^{\beta\,\hbar\,\omega}+1}{\left(e^{\beta\,\hbar\,\omega}-1\right)^2}\,A_1(N)+\langle \widehat{V}_R\rangle^2,
\end{eqnarray}\label{<VN>}
\end{subequations}
where 
\begin{eqnarray}
A_n(N)\equiv \sum\limits_{k=1}^{N}\frac{1}{\sqrt{n\,N^2}}\,\exp\left[-\frac{\pi\,(k-k_0)^2}{n\,N^2} \right].
\end{eqnarray}\label{A}
One can verify numerically that $A_n(N)$ is a real constant obeying $A_n(N)\leqslant 1$ for all $N$. Now we can implement the thermodynamic limit just by taking $N\to\infty$, with some additional care. For the high-temperature limit ($\beta\,\hbar\,\omega\to 0$), we see that these expectation values explode. However, this regime is supposed not to be suitably described by the master equation, which presumes weak coupling. For the low-temperature limit ($\beta\,\hbar\,\omega\to \infty$) on the other hand, a solution for the formulas \eqref{<VN>} is allowed to exist for $N\to\infty$, in agreement with the analysis carried out in Sec.\ref{thermo}. Finally, we note that in this case we have $\langle\widehat{V}_R^2\rangle\to\langle \widehat{V}_R\rangle^2$, indicating the disappearance of some types of correlations.

The structure of equation \eqref{ME} may be promptly identified to
the phase-destroying master equation \eqref{PDME}
commonly used in SID approaches. However, a rather important
conceptual difference must be noted in the (nonunitary) diffusion term, which
is in fact the one responsible for phase decoherence. In SID
approach, $\widehat{H}$ denotes the Hamiltonian of the system, whereas in
EID approach the decoherence of the system is induced by its
interaction with the reservoir. If on one hand these approaches
share the same mathematical structure, on the other they cannot be
conciliated physically.

The dependence of Eq.\eqref{ME} on $t$ occurs due to the nature of the coupling employed to model the reservoir. In fact, this is expected for models of reservoir inducing Gaussian decay in the quantum coherences, as can be concluded from the formulas \eqref{PDME}-\eqref{Psid}. A natural extension of our model in order to contemplate the exponential decay commonly observed in several contexts may be tried by assuming an adequate temporal dependence for the coupling $g_{0k}$. Research on this topic is in progress.

\section{Disentanglement induced by phase decoherence}
\label{disent}

Consider now the case in which the system of interest is composed
by two coupled oscillators. The Hamiltonian of the system is now
considered to be
\begin{eqnarray}
\widehat{H}_S = \hbar\,\omega_a\, \widehat{n}_a+ \hbar\,\omega_b \,\widehat{n}_b + \hbar \,g_{ab}\,\widehat{n}_a\, \widehat{n}_b.
\end{eqnarray}
Adopting the same algebraic procedure of Sec.\ref{decoherence}, we
obtain the characteristic function
\begin{eqnarray}
&C_{m_a,m_a' \atop m_b,m_b'}(t)=\prod\limits_{k=1}^N
\frac{1-e^{-\beta\,\hbar\,\omega_k}}{1-e^{-\beta\,\hbar\,\omega_k}\,
e^{-\imath\,[(m_a-m'_a)\, g_{ak}+(m_b-m'_b)\, g_{bk}]\,t} }&, \nonumber \\
\end{eqnarray}
which is associated to the two-boson state
\begin{eqnarray}
\widehat{\rho}_{S}(t)&=&\sum\limits_{m_a,m_a'=0 \atop m_b,m_b'=0 }^{\infty} \rho_{m_a,m_a' \atop m_b m_b'} \, C_{m_a,m_a' \atop m_b m_b'}(t)\,e^{-\imath(m_a-m_a')\,\omega_a t}\nonumber \\ &\times&\,e^{-\imath(m_b-m_b')\,\omega_b t}\,|m_am_b\rangle\langle m_a'm_b'|,\label{rhoS2t}
\end{eqnarray}
where
\begin{eqnarray}
\rho_{m_a,m_a' \atop m_b,m_b'}=\langle m_a m_b|\widehat{\rho}_{ab}(0)|m'_am'_b\rangle.
\end{eqnarray}
Also, it is possible to show that
\begin{eqnarray}
&|C_{m_a,m_a' \atop
m_b,m_b'}|^2=\prod\limits_{k=1}^N\left\{1+\frac{\sin^2\left[\frac{(m_a-m_a')
g_{ak} t+(m_b-m_b') g_{bk} t }{2} \right]
}{\sinh^2\left[\frac{\beta\,\hbar\,\omega}{2} \right]}
\right\}^{-1}& \nonumber \\&&
\end{eqnarray}
The analysis realized in the precedent sections concerning the short and long-time regimes can be also applied here with minimal changes. For the short-time
limit we obtain
\begin{eqnarray}
|C_{m_a,m_a' \atop m_b,m_b'}(t)|^2\cong e^{-t²/\tau_{ab}^2 },
\end{eqnarray}
where the decoherence time is now given by
\begin{eqnarray}
\tau_{ab}^{-2}=\sum\limits_{k=1}^N\left[\frac{g_{ak}\,(m_a-m_a')+g_{bk}\,(m_b-m_b')}
{2\,\sinh\left(\beta\,\hbar\,\omega/2\right)} \right]^2.
\end{eqnarray}
These expressions show how the decoherence process transforms
\begin{eqnarray}
\widehat{\rho}_{S}(0)&=&\sum\limits_{m_a,m_a'=0 \atop m_b,m_b'=0 }^{\infty} \rho_{m_a,m_a' \atop m_b m_b'} \,|m_am_b\rangle\langle m_a'm_b'|\label{rhoS20}
\end{eqnarray}
into
\begin{eqnarray}
\widehat{\rho}_{S}(t>\tau_{ab})&=&\sum\limits_{m_a,m_b=0}^{\infty} \rho_{m_a,m_a \atop m_b m_b} \,|m_am_b\rangle\langle m_a m_b|,\label{rhoS28}
\end{eqnarray}
which has no off-diagonal terms in the Fock basis. Here again,
additional care is required for the analysis of the critical
temperature regime (for $N$ finite), in which $|C|^2$ may assume finite values in the equilibrium, indicating the presence of surviving
entanglement. Also, as it occurred in the precedent
sections, for extremely low temperatures the degree of
entanglement may remain practically unchanged.

Let us now consider the simplest entangled initial state for the
system of two oscillators, namely
\begin{eqnarray}
|\psi_0\rangle=\frac{1}{\sqrt{2}}\Big(|nn\rangle+|mm\rangle\Big),
\end{eqnarray}
with $n \ne m$. Since the interaction does not couple different
number states, the above vector plays the role of a two-qubit
state coupled to the reservoir. In this case, we may follow the
procedure adopted in ref.\cite{emerson}, in which entanglement has
been evaluated by means of two-qubits measures, as for instance,
the negativity. This quantity has the following definition
\cite{TzuChieh03,Verstraete01}:
\begin{eqnarray}
{\mathcal N}[\widehat{\rho}(t)]=2\max\left\{0,-\lambda_{neg}(t)\right\},\label{nega}
\end{eqnarray}
where $\lambda_{neg}(t)$ is the negative eigenvalue of the
partially transposed density matrix in the joint space
$\mathcal{E}_{ab}$. Following \cite{emerson} we obtain
analytically:
\begin{eqnarray}
\lambda_{neg}(t)&=&-\frac{1}{2}\sqrt{C_{n,m \atop n,m}(t)\,C_{m,n \atop m,n}(t)}.
\label{lambda}
\end{eqnarray}
In the high-temperature and short-time regime we obtain
\begin{eqnarray}
\lambda_{neg}(t)\cong -\frac{1}{2}e^{(n-m)^2 t^2/\tau_{ab}^2},
\label{lambda1}
\end{eqnarray}
where
\begin{eqnarray}
\tau_{ab}=2\,\cal{G}_{ab}^{-1}\,\sinh\left(\beta\,\hbar\,\omega/2\right),
\end{eqnarray}
with $\cal{G}_{ab}^2\equiv\sum_k (g_{ak}+g_{bk})²/2$. Equations
\eqref{lambda1} and \eqref{nega} show how entanglement disappears
as the dynamics takes place, provided the temperature is higher than
the critical value given by Eq.\eqref{betaC} for the case of $N$ finite.  Also,
Eq.\eqref{lambda1} makes explicit that the disentanglement is
induced by the same mechanism that produces decoherence, namely,
the Gaussian decay in the characteristic function $C(t)$.

\section{Concluding remarks}\label{conclusions}

Inspired by a recent work \cite{fernando} we propose and theoretically investigate the properties of a reservoir composed by harmonic oscillators coupled to the system of interest by means of a cross-Kerr interaction. We have shown that this type of phase coupling is sufficient to guarantee decoherence without dissipation.

Since the analysis have been carried out analytically in virtue of the simplicity of the model, the main features of our cross-Kerr reservoir could be clearly identified. For the regime of short times and finite temperatures, a state-independent decoherence time has been estimated and a Gaussian decay for quantum coherences has been demonstrated through the analysis of the characteristic function $C_{m,m'}(t)$. 

Concerning the coupling, we have concluded that the precise profile of the spectral distribution is relevant only in order to guarantee the irreversibility of the dynamics, not being important to decide whether the decoherence may occur or not. In fact, the mechanism responsible for the decoherence has been shown to be the Gaussian decay (derived from the features of both the coupling and the thermal state), which is always present in the characteristic function for short times independently either on the regime of temperature or the number of oscillators in the reservoir. 

By analyzing the long-time dynamics for finite reservoirs we have detected regimes of partial decoherence, in which the occurrence of surviving off-diagonal terms in the quantum state is expected. In this case, it was possible to identify a critical temperature beyond which decoherence is expected to be total. In the thermodynamic limit ($N\to\infty$), however, the regime of partial decoherence no longer exists since $T_{crit}\to 0$.

We have also derived a phase-destroying master equation without statistical assumptions in the weak coupling approximation. The thermodynamic limit has been derived for the regime of low temperatures, indicating the occurrence of decohering dynamics under these conditions. These results suggest that the EID approach is able to produce most of the SID results with no need for additional postulates in quantum theory.

Last but not least, we have verified a Gaussian decay of the entanglement associated to the a bipartite system coupled to our cross-Kerr reservoir. The result, which has been obtained for arbitrary $T$ and $N$, illustrates the role of disentanglement being induced by the same mechanism responsible for the decoherence.

\acknowledgments{
The authors would like to thank M.H.Y. Moussa and F.L. Semião for helpful suggestions and fruitful discussions. KF and ESC acknowledge support from CNPq under projects $\#300651/85-6$ (KF - partial) and $\#140243/2001-1$ (ESC).}

\newpage

\end{document}